\documentclass[twocolumn,aps,showpacs,pra,preprintnumbers,bibnotes10pt,superscriptaddress]{revtex4}

\usepackage{graphicx}
\usepackage{dcolumn}
\usepackage{epsfig}
\usepackage{bm}
\usepackage{array}
\usepackage[frenchb]{babel}

\begin{document}

\title{Direct evaporative cooling of $^{39}$K atoms to Bose-Einstein condensation}
\author{M. Landini}
\email{landini@lens.unifi.it}\affiliation{CNR Istituto Nazionele Ottica, 50019 Sesto Fiorentino, Italy}
\affiliation{LENS and Dipartimento di Fisica e Astronomia, Universit\'a di Firenze, 50019 Sesto Fiorentino, Italy}
\author{S. Roy}
\affiliation{LENS and Dipartimento di Fisica e Astronomia, Universit\'a di Firenze, 50019 Sesto Fiorentino, Italy}
\author{G. Roati}
\affiliation{CNR Istituto Nazionele Ottica, 50019 Sesto Fiorentino, Italy}
\affiliation{LENS and Dipartimento di Fisica e Astronomia, Universit\'a di Firenze, 50019 Sesto Fiorentino, Italy}
\author{A. Simoni}
\affiliation{Institut de Physique de Rennes,  UMR 6251 CNRS-Universit\'e de Rennes 1, 35042 Rennes Cedex, France}
\author{M. Inguscio}
\affiliation{LENS and Dipartimento di Fisica e Astronomia, Universit\'a di Firenze, 50019 Sesto Fiorentino, Italy}
\affiliation{CNR Istituto Nazionele Ottica, 50019 Sesto Fiorentino, Italy}
\affiliation{INFN, Sezione di Firenze, 50019 Sesto Fiorentino, Italy}
\author{G. Modugno}
\affiliation{LENS and Dipartimento di Fisica e Astronomia, Universit\'a di Firenze, 50019 Sesto Fiorentino, Italy}
\affiliation{INFN, Sezione di Firenze, 50019 Sesto Fiorentino, Italy}
\author{M. Fattori}
\affiliation{CNR Istituto Nazionele Ottica, 50019 Sesto Fiorentino, Italy}
\affiliation{LENS and Dipartimento di Fisica e Astronomia, Universit\'a di Firenze, 50019 Sesto Fiorentino, Italy}
\affiliation{INFN, Sezione di Firenze, 50019 Sesto Fiorentino, Italy}

\date{\today}

\begin{abstract}
We report the realization of Bose-Einstein condensate of $^{39}$K atoms without the aid of an additional atomic coolant. Our route to Bose-Einstein condensation comprises sub-Doppler laser cooling of large atomic clouds with more than $10^{10}$ atoms and evaporative cooling in an optical dipole trap where the collisional cross section can be increased using magnetic Feshbach resonances. Large condensates with almost $10^6$ atoms can be produced in less than 15 seconds. Our achievements eliminate the need for sympathetic cooling with Rb atoms which was the usual route implemented till date due to the unfavorable collisional property of $^{39}$K. Our findings simplify the experimental set-up for producing Bose-Einstein condensates of $^{39}$K atoms with tunable interactions, which have a wide variety of promising applications including atom-interferometry to studies on the interplay of disorder and interactions in quantum gases.
\end{abstract}


\maketitle

\section{Introduction}

In recent years, many interesting observations have been made possible by the use of Bose Einstein condensates with tunable interactions \cite{Inouye}. Magnetic Feshbach resonances have been used to study the formation of bright solitons at small and negative scattering lengths \cite{Khaykovich, Strecker, Cornish}. Tuning the two body interactions, many collective phenomena have been explored using degenerate quantum gases loaded in optical lattices \cite{HallerF, HallerS}. The almost complete cancellation of the interactions between the atoms has allowed the sensitivity of BEC based atom interferometers to be enhanced \cite{Gustavsson, FattoriF, FattoriS}. A fine control of the collisional properties of a gas has offered new possibilities in the study of the interplay between disorder and interaction in matter waves \cite{Roati,Deissler,Lucioni,Hulet}. Finally, ultracold gases with tunable interactions have led to the first observation of an Efimov spectrum for resonantly interacting particles, contributing to the experimental validation of important few-body physics theoretical predictions \cite{Kraemer,Ferlaino}.

Although magnetic Feshbach resonances are present in all alkali species, only a few bosonic isotopes have sufficiently broad resonances to enable an easy manipulation of the scattering length over several orders of magnitude. A list of systems includes $^7$Li \cite{HuletLi}, $^{133}$Cs \cite{Innsbruck} and $^{85}$Rb \cite{Cornell}. Very recently also $^{39}$K has proven to be a good candidate for experiments where a high degree of interaction tunability is needed \cite{roati, Hadzibabic2}, thanks to a broad magnetic Feshbach resonance with a 52 G width at a relatively low magnetic field of 402 G. The main difficulty in using $^{39}$K atoms is its collisional properties at zero magnetic field. Contrary to $^{41}$K \cite{inouye}, a background scattering length of -33 a$_0$ compromises the stability of a $^{39}$K BEC and, more importantly, gives rise to a Ramsauer minimum in the collisional cross section at 400 $\mu$K that makes a direct evaporation to condensation quite unfavorable \cite{Luigi}. This hurdle was overcome for the first time by sympathetic cooling with $^{87}$Rb atoms \cite{roati}, a technique that has been applied in the past also for the other potassium isotopes \cite{sympathetic,40K}. However, this method requires an additional species to cool the atomic sample of $^{39}$K, with the associated complications and instabilities, and generally a preparation time of the order of one minute is needed to reach condensation \cite{roati, Hadzibabic}. This has made the procedure to condense $^{39}$K more complicated as compared to other alkali atoms with more favorable collisional properties.

In this paper, we report on the realization of Bose-Einstein condensates of $^{39}$K atoms without the aid of sympathetic cooling with other atomic species. We describe the various cooling steps performed in a new apparatus, which was optimized for this purpose \cite{manu}. Using an optimized 2D$^+$ MOT - 3D MOT configuration we are able to collect in a few seconds almost 3 $\cdot$ 10$^{10}$ atoms. By laser cooling the cloud to sub-Doppler temperatures as low as 25 $\mu$K, and trapping the atoms in a strongly compressed magnetic quadrupole trap, we are able to significantly increase the phase space density. This is however still not enough to activate an efficient evaporative cooling in the magnetic trap. We could however transfer about $3 \cdot 10^7$ atoms in a deep optical dipole trap, where we can perform a very efficient evaporation by a proper tuning of the scattering length. Almost pure condensates of up to $8 \cdot 10^5$ atoms can be produced with a total sequence duration of less than 15 seconds, a factor of four faster than previously reported. Thanks to the single species operation, the stability of the experimental apparatus is also increased.

The article is divided into eight sections. In section II we give a general description of the experimental setup. In section III-VII we provide details of the different cooling stages, i.e. 2D$^+$ MOT (Sect. III), 3D MOT (Sect. IV), magnetic transport (Sect. V), loading of dipole trap (Sect. VI) and evaporative cooling towards quantum degeneracy (Sect. VII). In section VIII, we summarize our major achievements and list possible future improvements.

\section{Experimental set-up}
The experimental set-up consists of three chambers connected by low conductivity tubes to achieve proper differential vacuum levels (see Fig. 1). In the first chamber, a high flux of slow atoms is produced with an optimized 2D$^{+}$MOT. Atoms are then collected into a 3D MOT in the second cell. Laser-cooled atoms are then trapped magnetically and thereafter transported to the science cell using a moving magnetic transport system. Here evaporative cooling techniques have been applied to achieve quantum degenerate samples. A three-chamber apparatus is needed in order to fulfill the requirements of a fast and large atom number loading and a wide optical access in the final science glass cell.

\begin{figure}[ht] \label{K39}
\begin{center}
\includegraphics[width=\columnwidth] {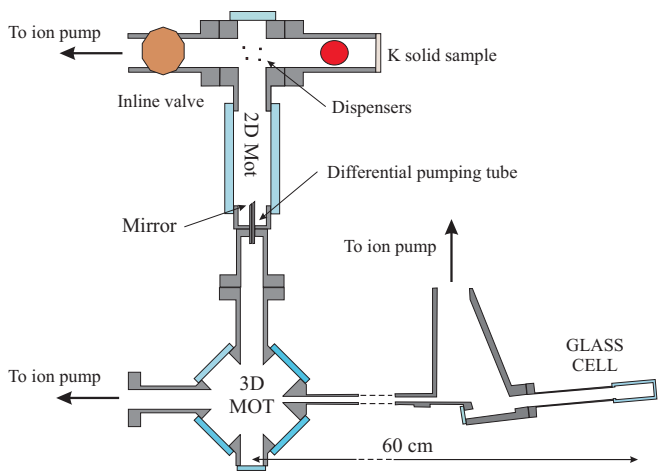}
\end{center}
\caption{Schematic of the vacuum system. Note that the windows are directly glued to the 2D MOT and 3D MOT chambers made of medical titanium.}
\end{figure}

\section{2D$^+$ MOT}
In the first chamber, we implement a 2D$^+$ MOT similar to the one described in \cite{Catani}. The narrow hyperfine structure of $^{39}$K requires the use of equal power in the cooling and repumper lasers. The cooling light is red-detuned by about 6 $\Gamma$, while the repumping light is kept on resonance. A large total power of 400 mW has been used to increase the velocity capture range and the atomic flux. We employ two transverse beams with 9 mm and 44 mm vertical and horizontal waist, respectively. The two-dimensional MOT is operated in the 2D$^+$ configuration, by employing two additional laser beams along the atomic beam direction. The "push" beam, propagating in the direction of the atomic flux and the "retarding" beam, propagating in the opposite direction. These beams have a waist of 5 mm and the same frequencies as the transverse ones. The retarding beam is reflected by a mirror inside the vacuum chamber, at $45 \degres$ with respect to the atomic flux direction. The metal substrate of the mirror was gold-coated for high reflectivity. Furthermore, an extra MgF$_2$ coating was realized on top of the gold one to avoid deposition of potassium.
The mirror has a 1.5 mm diameter hole, 30 mm long to reduce the conductivity towards the 3D MOT chamber. Adjusting the power in the "push" and "retarding" beams within the order of a few mW, we are able to optimize the atomic beam average velocity to maximize the 3D-MOT loading. Optimum performances were achieved with a flux of 2 x 10$^{10}$ atoms/s, with an average velocity of 25 m/s.

The potassium vapor pressure can be controlled using commercial getters assembled very close to the cooling region, or making use of a solid potassium sample. After heating the sample to $100 \degres$C for a few days we could have optimum operation for several months keeping the whole vacuum chamber at room temperature. This is surprising considering that potassium vapor pressure at 300 K has a rather low value of $10^{-8}$mbar. In order to slow down the depletion of potassium from the 2D-MOT region, we have limited the pumping speed of the ion pump by almost closing an inline valve between the pump and the 2D-MOT cell.

\section{3D-MOT}
In the second chamber, we slow down the atomic beam, trap and cool the potassium atoms using a 3D MOT. The operation is divided into three different sequences, identified as loading, compression and molasses. Optimum operation parameters have been listed in Table \ref{parameters}. A total power of 400 mW, available for both cooling and repumper light, is split into six independent beams with a waist of 17.5 mm. During the MOT loading, we estimate a velocity capture range of the order of 50 m/s, well above the average velocity of the atomic beam. Clouds of almost $3 \cdot 10^{10}$ atoms at a temperature of 2 mK can be captured in 2-3 seconds.
During the compressed MOT phase, we strongly decrease the repumper intensity and increase the cooling light detuning to suppress light-assisted collisions that prevents the atoms to get too close. This step allows an increase of the atomic density by a factor of 10.
\begin{table*}[ht]
\centering 
\begin{tabular}{c c c c c c c c c c} 
\hline
Sequences & $\delta_C/\Gamma$ & $\delta_R/\Gamma$ & $I_{tot}/I_{S}$ & $I_R/I_C$ & $\gamma (G/cm)$ & $T(\mu K)$ & N & n $(atoms/cm^3)$ & $\rho$ \\ [1ex] 
\hline 
Loading (3 s long) & -3 & -3.3 & 35 & 0.8 & 11 & 2000 & 3x$10^{10}$ & 1.8 x $10^{10}$ & 4.6 x $10^{-9}$\\ 
C-MOT (values kept for 5 ms) & -3.7 & 0 & 35 & 0.5 & 15 & 2000 & 1.8 x $10^{10}$ & 1.7 x $10^{11}$ & 4.6 x $10^{-9}$ \\
C-MOT (ramp 10 ms long) & -3.7 to -6.2 & 0 & 35 & 0.5 to 0.02 & 15 & 2000 & 1.8 x $10^{10}$ & 1.7 x $10^{11}$ & 4.6 x $10^{-9}$ \\
Molasses initial values & -0.7 & -2.7 & 18 & 0.01 & 0 & - & - & - &\\
Molasses ramp (10 ms long) & -0.7 to -2.5 & -2.7 & 18 to 1 & 0.01 & 0 & 25 & 1.65 x $10^{10}$ & 8.1 x $10^{10}$ & 1.5 x$10^{-5}$ \\
Quadrupole trap loading & / & / & 0 & / & 30 & 55 & 3.8 x $10^{9}$ & 1.8 x $10^{11}$ & 1 x$10^{-5}$ \\
Science cell quadrupole trap & / & / & 0 & / & 270 & 250 & $10^{9}$ & 4 x $10^{11}$ & 2.3 x$10^{-6}$ \\ [1ex] 
\hline 
\end{tabular}
\caption{Experimental parameters implemented during the various experimental sequences. For each stage, we report the optimum temperature $T$, atom number $N$, spatial atom number density $n$ and achieved phase-space density $\rho$. $\Gamma$ is the natural linewidth of the excited state, $I_S$ is the saturation intensity, $\gamma$ is the magnetic field gradient.}
\label{parameters} 
\end{table*}
The molasses sequence is the most critical and the most important part of the atomic cloud preparation. Extensive description of sub-Doppler cooling of $^{39}$K can be found in \cite{Landini}. The main idea is to tune the cooling light frequency to the red of the strong $\vert F=2 \rangle \to \vert F'=3 \rangle$ D2 optical transition to exploit sub-Doppler forces, but sufficiently blue-detuned with respect to the $\vert F=2 \rangle \to \vert F'=2 \rangle$ one, to avoid negative friction forces. Note that the small separation of 3$\Gamma$ between the excited states $F'=3$ and $F'=2$, where $\Gamma$ is the natural linewidth, might lead to high scattering rates and strong heating. To reduce such effect and lower the minimum temperature achievable, we ramp down the cooling light intensity and keep a very low ratio (about 1/100) between the repumper and the cooling light intensities. In this way, we leave most of the atoms in the F=1 state and minimize the amount of light rescattered by the atoms during their cooling.

\section{Magnetic transport}
At the end of the molasses phase, we optically pump the atoms in a state that can be magnetically trapped. The $\vert F=2, m_F=2 \rangle$ state was our first choice because it is the low field seeking state which allows maximum magnetic confinement. Unfortunately, due to the narrow hyperfine level structure of the excited states, it is not a pure dark state for $\sigma^+$ polarized light tuned on the $\vert F=2 \rangle \to \vert F'=2 \rangle$ transition. As a consequence, at our best we can polarize only 70 \% of the atoms with an unavoidable heating of the cloud to 60 $\mu$K. In order to get a fully polarized sample, we implement a much simpler procedure. At the end of the molasses we switch off the repumper and transfer all the atoms in the F=1 manifold. Applying a quadrupole field of 30 Gauss/cm instantly, we lose approximately two thirds of the atoms in the $\vert F=1, m_F=0,+1 \rangle$ states but we get a fully pure sample in $\vert F=1, m_F=-1 \rangle$ trapped magnetically. Note from Table \ref{parameters} that avoiding heating, during the transfer from the molasses to the magnetic quadrupole trap, we have only an about 30\% reduction of the phase space density.
In order to transfer the atoms from the 3D MOT chamber to the glass cell we ramp up the current in a pair of coils mounted in an anti-Helmoltz configuration on a moving cart. A total magnetic field gradient of 165 G/cm allows one to transport the atoms over a distance of 540 mm in 2 seconds with negligible heating. However, a few seconds lifetime in the tubes connecting the 3D MOT chamber with the glass cell and a cloud size comparable with the tube's cross-section diameter cause a factor of 4 reduction in the atom number and hence the phase-space density. In the science cell we get about 10$^9$ atoms at a temperature of $250 \mu$K after an adiabatic ramp of the magnetic quadrupole field up to 270 G/cm (see Table \ref{parameters}).
\begin{figure}[ht]
\begin{center}
\includegraphics[width=\columnwidth] {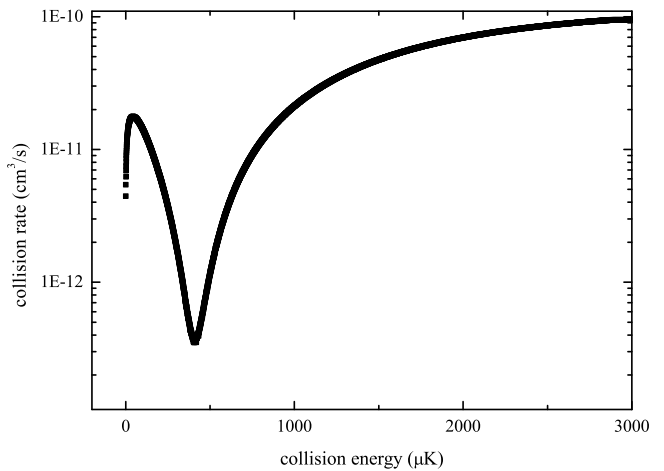}
\end{center}
\caption{Numerically calculated elastic collision rate $\sigma v$, where $\sigma$ is the elastic cross-section and $v$ the relative velocity of the atom pair, as a function of the collision energy in the center-of-mass frame. The calculation is performed for atoms in the $| F=1,m_F=-1 \rangle$ hyperfine state in a vanishing magnetic field, and it includes $s$ and $d$ partial waves. Collision model parameters are as in \cite{NJP}.} \label{2DMOT}
\label{efficiency}
\end{figure}

\section{Loading of the optical dipole trap}
Due to a Ramsauer minimum in the elastic cross-section at 400 $\mu$K (see Fig. 2) the average collisional rate for the atoms trapped in the quadrupole is below 1 Hz and consequently too low for an efficient evaporation. In order to overcome such obstacle, we load the atoms into a far-detuned dipole trap in order to increase the collisional rate exploiting a Feshbach resonance. We implement a single beam derived from a 100 W IPG-Photonics Ytterbium multimode fibre laser (Model YLR-100-LP-AC) tightly focused to a waist $w$=25 $\mu$m, aligned close to but not exactly at the centre of the quadrupole to prevent Majorana spin flips. The laser is switched on abruptly and after an optimum loading time, the magnetic trap is switched off instantaneously. A measurement of the optically trapped atoms as a function of the loading time is reported in Fig. 3 for different laser beam powers. In the following we will try to explain the optimum loading behavior achieved for 27 Watts of trapping laser power.

\begin{figure}[ht]
\begin{center}
\includegraphics[width=\columnwidth] {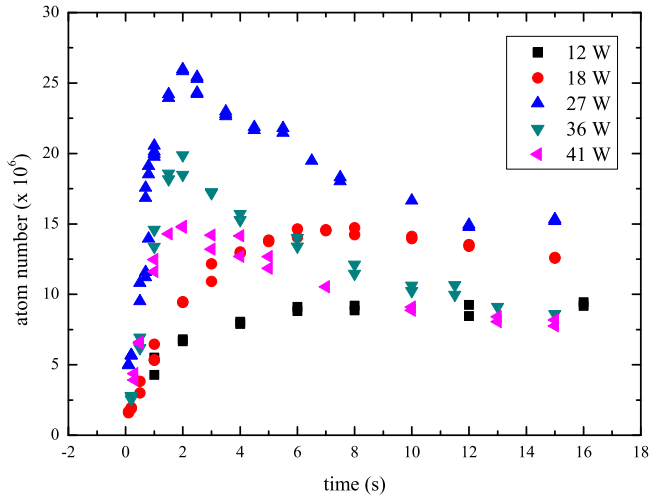}
\end{center}
\caption{Atom number trapped from the quadrupole into the dipole trap as a function of the loading time. The measurement is performed for different values of the laser power.} \label{2DMOT}
\label{efficiency}
\end{figure}

For small trapping times, the number of atoms is $\sim 10^6$. This corresponds to $n_Q (\pi w^2 l_Q)$, i.e. the number of atoms occupying the dipole trap volume. Here $n_Q \sim 5 \cdot 10^{11}$ cm$^{-3}$ is the atom number density in the quadrupole trap, $l_Q=(N_{a}/n_Q)^{1/3} \sim 1$ mm is the average size of the quadrupole trap and $N_{a} \sim 10^9$ is the number of atoms initially trapped in the quadrupole trap. The loading rate at short times can be estimated using an analysis similar to the one presented in \cite{Pillet}. With simple geometrical arguments, we can state that only $N_a (w/l_Q) \sim 10^7$ atoms have initially a trajectory passing through the dipole trap. At a generic time, only a fraction $t_{transit}/t_{osc}$ of these atoms are in the dipole trap, where $t_{transit}$ is the transit time through the dipole trap and $t_{osc}$ is the oscillation time in the quadrupole trap. Note that $t_{transit} \sim 10^{-4}$s can be estimated calculating the inverse of the radial trapping frequency of the dipole trap, while $t_{osc}$ is of the order of $l_Q / \langle v \rangle \sim 3$ ms, i.e. the size of the magnetically trapped cloud divided by the average thermal velocity of the atoms in the quadrupole trap. In addition, during a small time interval $\Delta t$, only $\Delta t /\tau$ atoms can collide, where $1/ \tau$ is the collisional rate in the dipole trap. This rate can be estimated multiplying the initial density, that is of the order of the atomic density in the quadrupole trap times the elastic cross section at the average collisional energy in the dipole trap, $\sim$ few mK (see Fig. 2). Note that this collisional rate is much larger than in the bare quadrupole trap because the average collisional energy in the dipole trap is of the order of the trap depth and so well above the Ramsauer minimum. The numerical value for the loading rate can be estimated from the expression $N_a (w/l_Q) (t_{transit}/t_{osc}) n_Q \langle \sigma v \rangle $ where $\langle \sigma v \rangle \sim 5 \cdot 10^{-11}$ cm$^3$/s. Using the numerical values reported above, we get a loading rate of the order of $10^7$ atoms/s consistent with our measurement. At longer times, we can identify two competing effects. The first one is a reduction of the atom number with the decay law of the density given by $n'(t)=-K_2 n^2(t)$, where $K_2$ is the two-body loss coefficient which was measured to be $\sim 10^{-13}$ cm$^3$/s, for an intensity of the dipole trap laser of 3 MW/cm$^2$. For our typical atom number density of $\sim$10$^{13}$ atoms/cm$^3$, achieved after the dipole trap loading, this corresponds to an initial loss rate of $\sim$ 1 Hz. We have attributed these two-body losses to light-assisted inelastic collisions, enhanced by the multimode spectrum of our laser \cite{Birkl, Bagnato, julienne}. The second is a residual loading from the quadrupole. After all the atoms initially passing through the dipole trap has been captured ($N_a w/l_Q \sim 10^7$), it is necessary to wait an "ergodicity time" required to modify the trajectories of the atoms in the quadrupole. However, this time is a few seconds long, comparable with the thermalization time in the quadrupole trap. The optimum laser power of 27 W balances these two contributions. Higher powers enhance the inelastic collision rate without a net increase of the elastic one (see the flat slope at high collision energies in Fig. 2). As a consequence, for powers larger than 27 W, at short times we observe similar loading rates, but at longer times we experience stronger losses (see Fig. 3). On the other hand, decreasing the power reduces the collision energy in the dipole trap and consequently the elastic cross section and the loading rate. As we can see from Fig. 2, the collision rate decreases very rapidly below 1 mK collision energy. This is confirmed experimentally by a very fast reduction of the loading rate for decreasing powers. Note that a 10 W trapping laser power corresponds to 700 $\mu$K trap depth.
\begin{figure}[ht]
\begin{center}
\includegraphics[width=\columnwidth] {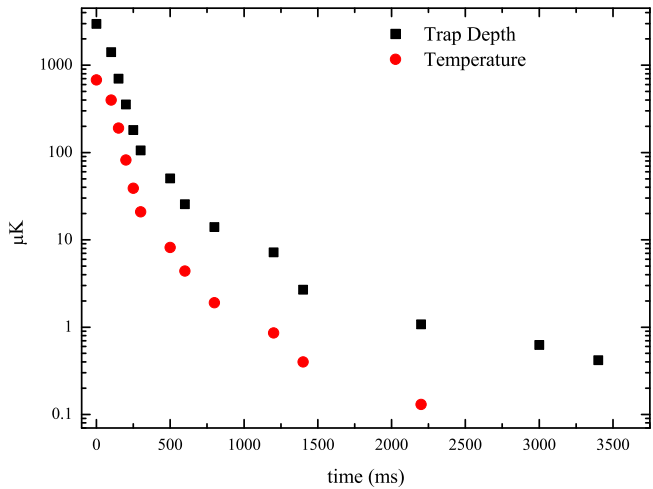}
\end{center}
\caption{Optimized variation of the dipole trap power during the evaporative cooling. Solid squares and solid circles represent the estimated trap depth and the measured temperature of the atoms respectively.} \label{ev}
\label{efficiency}
\end{figure}

\section{Evaporative cooling to Bose Einstein condensation}

The evaporative cooling is performed by reducing the IPG laser power from 27 W to about 5 mW in 3.5 seconds by means of an acousto-optical modulator (see Fig. 4). The radial and longitudinal trapping frequencies are estimated to be initially 5 kHz and 50 Hz respectively.
During the initial part of the evaporative cooling, the scattering length of the atoms is appropriately tuned using a low field Feshbach resonance at 33.6 G for atoms in the m$_F$=-1 state. Note that, although the initial lower atomic density would in principle allow one to use a large scattering length, enhanced light-assisted collisional losses close to the resonance strongly limit its optimum value to 16 a$_0$, obtained at 70 Gauss. At this field, we measure a $K_2$ value of $\sim$ 4$\times$10$^{-14}$ cm$^3$/s. After 100 ms, the lower intensity of the laser makes the tuning of the scattering length to a larger value of 88 a$_0$ more effective.
After one second, we perform a Landau-Zener RF sweep to transfer the cloud to a given target internal state and perform the last 2.5 seconds of evaporative cooling in proximity of a Feshbach resonance. We could reach condensation in each one of the 3 magnetic sub-levels $m_F$ of the hyperfine ground state $F=1$ manifold, on a total of 5 different Feshbach resonances located at the $B{_F}$ magnetic field ($(m_F, B_F)=(1, 402.4$ G $), (0, 471$ G $), (-1, 33.6$ G $), (-1, 162.3$ G $), (-1, 560.7$ G $)$) \cite{NJP}, with similar atom numbers in the condensates. In Fig. 6, we show the evaporative cooling efficiency for atoms in the $F=1, m_F=+1$ state, tuning the interparticle interaction with the Feshbach resonance at 402.4 Gauss. Condensates with up to $8 \times 10^5$ atoms can be produced in a trap with radial and longitudinal trapping frequencies of 50 Hz and 2 Hz, where the gravitational force is partially compensated with a vertical magnetic field gradient. Note that the longitudinal confinement is generated by a weak magnetic curvature originating from non-perfect Helmholtz configuration of our Feshbach coils.
\\
\begin{figure}[ht]
\begin{center}
\includegraphics[width=\columnwidth] {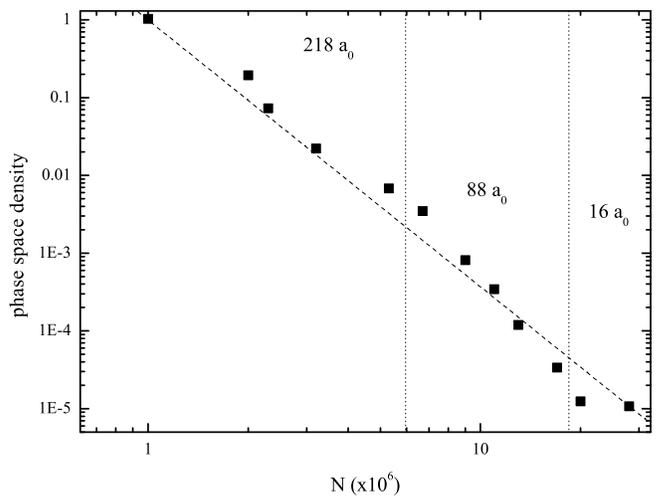}
\end{center}
\caption{Phase-space density vs atom number during the evaporative cooling for the realization of condensate in the $m_F=1$ state close to the 402.4 G Feshbach resonance. The vertical dotted lines are placed where the s-wave scattering length is changed. The dashed line connecting the initial and final points corresponds to an over-all evaporative cooling efficiency of 3.4. Note that throughout the lowering of the trapping beam power, we find more efficient evaporative cooling by tuning the scattering length to larger values.} \label{eff}
\label{efficiency}
\end{figure}
\\
In Fig.\ref{eff}, the achieved phase-space density vs atom number is reported for the optimized evaporation ramp.
\begin{figure}[ht]
\begin{center}
\includegraphics[width=\columnwidth] {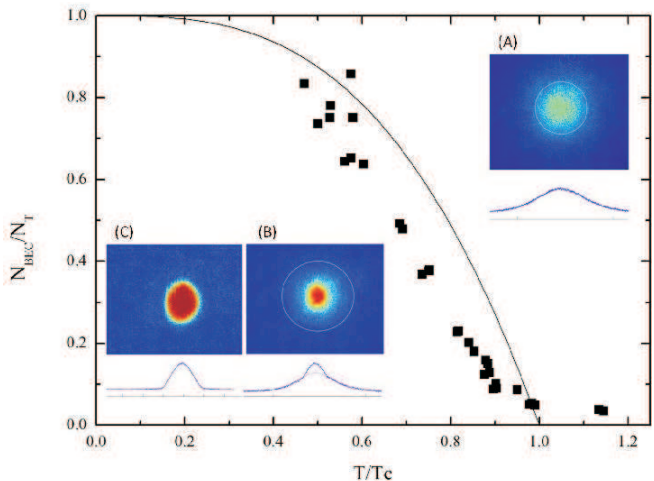}
\end{center}
\caption{Condensed fraction vs temperature around the BEC phase transition point. The theoretical curve is $1-(T/T_c)^3$ and includes finite size and interaction shifts of the critical temperature \cite{Stringari}. For condensed fractions above 80\%, the temperature was not measurable due to the low signal in the thermal tails of the distribution. In the insets, we show the absorption images and the atomic density profiles of the cloud above (A), around (B) and below (C) the condensation critical temperature} \label{tc}
\label{efficiency}
\end{figure}
For completeness, in Fig.\ref{tc} we report the measurement of the condensed fraction and images of the atomic density profiles close to the BEC phase transition point. The measurements are consistent with the expected critical temperature of $\sim 150$ nK.

\section{Conclusion and outlook}
We have demonstrated the realization of Bose-Einstein condensates of $^{39}$K atoms without implementing any additional atomic coolant. The efficiency of the evaporative cooling in an optical dipole trap was enhanced by tuning the two-body s-wave scattering length using magnetic Feshbach resonances of the lower hyperfine ground states of $^{39}$K. Pure condensates with up to $8 \times 10^{5}$ atoms could be produced in less than 15 seconds, a factor of four faster than previously reported. The elimination of the need for sympathetic cooling with other atomic species greatly simplifies the experimental setup and procedure to obtain degenerate quantum gas of bosonic $^{39}$K atoms with tunable interactions, useful for a variety of applications like quantum interferometry and quantum simulation with cold atoms \cite{naturephys}. Further improvements in the condensation procedure might be achieved in the future by loading the atoms in the dipole trap directly from the molasses or/and implementing single frequency high power lasers to suppress light-assisted collisions.

\section{Acknowlegments}
We gratefully acknowledge contributions from L. Carcagn\'i and J. Catani. We thank all our colleagues of the Quantum Degenerate Group at LENS for inspiring discussions.
This work was supported by ERC (AISENS starting grant and DISQUA advanced grant), INFN (MICRA Collaboration), by EU (IP AQUTE, QIBEC) and by CNR (EuroQUASAR program).

\end{document}